\documentstyle[12pt]{article}
\newcommand{\be}{\begin{equation}}
\newcommand{\ee}{\end{equation}}
\begin{document}
\begin{center} 
{\Large\bf Solid-State Nuclear Spin Quantum Computer Based on Magnetic 
Resonance Force Microscopy}\\ \ \\
G.P. Berman$^1$, G.D. Doolen$^1$, P.C. Hammel$^2$, and V.I. Tsifrinovich$^3$
\end{center}
$^1$Theoretical Division and CNLS, Los Alamos National Laboratory, \\
Los Alamos, New Mexico 87545\\
$^2$ MST-10, Los Alamos National Laboratory, MS K764, Los Alamos NM 87545\\
$^3$Department of Physics, Polytechnic University,
Six Metrotech Center, Brooklyn NY 11201\\ \ \\
\begin{center}
{\bf ABSTRACT}
\end{center}
We propose a nuclear spin quantum computer based on magnetic resonance 
force microscopy (MRFM). It is shown that an MRFM single-electron spin 
measurement provides three essetial requirements for quantum computation 
in solids: (a) preparation of the ground state, (b) one- and two- qubit 
quantum logic gates, and (c) a measurement of the final state. The 
proposed quantum computer can operate at temperatures up to 1K. 
\newpage
\quad\\
{\bf I. Introduction}\\ \ \\
It is well-known that a quantum computer can be implemented in a chain of two-level quantum atoms connected by weak interactions \cite{lloyd93}. One-qubit rotations and two-qubit quantum logic gates can be implemented in this chain using resonant pulses which induce transitions between the energy levels of the system. The ``natural'' implementation of this idea is an Ising spin quantum computer which contains a chain of 1/2 spins placed in a permanent magnetic field and interacting through a weak Ising interaction \cite{ber94}.
 It has been realized that quantum computation is possible even for an ensemble of Ising spin chains at temperatures which are much higher than
the energy spacing between two stationary states of a spin \cite{m1}-\cite{m3}. High resolution liquid state NMR quantum computing has been extensively developed for a small number of spins \cite{m4}-\cite{m6}. 

Magnetic resonance force microscopy (MRFM) has matured over the past few years. It promises atomic scale detection of both electron and nuclear spins \cite{i1}-\cite{i2}. MRFM could provide the crucial step from  liquid quantum computer to the solid state quantum computer, which has the potential to incorporate a large number of qubits.

In this paper, we propose a nuclear spin quantum computer based on the use of MRFM. In Sec. 2, we describe  standard MRFM techniques and the method of a single-spin measurement suggested in \cite{bt1}. This method employs a periodic sequence of short resonant pulses instead of the continuous excitation used in a standard MRFM. In Sec. 3, we propose a strategy for detecting a nuclear spin state via the electron spin transition in a paramagnetic atom. This approach utilizes the hyperfine splitting of the electron spin resonance.
In Sec. 4, we describe an implementation of a MRFM nuclear spin quantum computer in a chain of impurity paramagnetic atoms in a diamagnetic host. We have shown that this quantum computer can provide the three essential requirements  for quantum computation: (a)  preparation of the ground state (the initial polarization of nuclear spins), (b) quantum logic gates,  and (c) a final measurement of the quantum states.
We also discuss a nuclear spin implementation of quantum computation based on a MRFM which does not use a single-spin detection.\\ \ \\
{\bf II. Single Spin Measurement}\\ \ \\
First, we shall discuss an important issue: How can one detect a single-spin using an MRFM? The main part of the MRFM is the cantilever. (See Fig. 1.) One end of the cantilever is free to vibrate.  Its extremely weak mechanical oscillations can be detected by optical methods. The second important part of the MRFM is a small ferromagnetic particle, $F$, which produces a non-uniform magnetic field, $B_F$, in the sample, $S$. The remaining parts are standard for NMR, ESR and FMR techniques: a uniform magnetic field, $B_0$, and a resonant radio-frequency ({\it rf}) field, $B_1$. The non-uniform magnetic field, $B_F$, produced by the ferromagnetic probe changes the magnetic field in the sample in such a way that only selected spins are resonant with the {\it rf} field. Changes in the orientation of selected spins under the action of the resonant {\it rf} field influence the magnetic force between the ferromagnetic particle and the sample, causing oscillations of the cantilever. So far, experimenters usually have to put the sample on the cantilever. 
For MRFM imaging, however, it would be very important to place not a sample but a ferromagnetic particle on the cantilever. This is an experimental challenge. We will assume, for definiteness, that the ferromagnetic particle can be placed on the cantilever, as shown in Fig. 1.

At present, MRFM utilizes the driven resonant oscillations of a cantilever. For example, in experiments with the ferromagnetic resonance \cite{zhw} with yttrium garnet film, both the magnetic field, $B_0$, and the {\it rf} field, $B_1$, were modulated at two frequencies whose difference (or sum) equals the cantilever's frequency, $f_c$ (anharmonic modulation). In \cite{zhw}, $B_0\sim 10^{-2}$T,
$B_1\sim 10^{-4}$T, the modulation frequency of the magnetic field, $f_{0m}=36.01$kHz, the modulation frequency of the
{\it rf} field was $f_{1m}=41.27$kHz, and the resonant frequency of the cantilever was $f_c=5.26$kHz.

In experiments with the proton magnetic resonance in ammonium nitrate \cite{rzh}, the modulation scheme was based on cyclic adiabatic inversion. The frequency of the {\it rf} field was modulated, so that the effective magnetic field in the rotating frame adiabatically changed its direction. The nuclear magnetization followed the effective field. As a result, the longitudinal (z-) component of the nuclear magnetic moment oscillated, producing a cyclic force on the cantilever. In this experiment, the permanent magnetic field, $B_0$ was $1.95$T, the amplitude of the {\it rf} field, $B_1$ was $\sim 10^{-3}$T, the frequency of the {\it rf} field was $\sim 100$MHz, and the modulation frequency equaled  the cantilever's resonant frequency, $f_c=1.4$kHz.

In our proposed quantum computer discussed in Secs 3 and 4, we use an electron single-spin measurement to determine a nuclear spin state. We should note that a single-spin measurement is quite different from the measurement of a macroscopic (classical) magnetic moment. First, we shall discuss a single-electron
 spin measurement using a static displacement of the  cantilever.
If the electron magnetic moment points in the positive $z$-direction (state $|0_e\rangle$) the magnetic force on the electron, $F_z=\mu_B\partial B_z/\partial z$, is negative.  In this case, the cantilever is attracted to the sample, $S$. Suppose, that this deflection from equilibrium of the cantilever's tip can be detected. Then the deflection of the cantilever's tip in the positive $z$-direction will indicate that the electron spin is in the ground state, $|0_e\rangle$. Similarly, a negative displacement of the cantilever indicates that the electron is in its  excited state, $|1_e\rangle$. Then, this cantilever is a classical measuring device: If the electron was initially in a superpositional state, it will jump to one of the states, $|0_e\rangle$ or $|1_e\rangle$, due to the interaction with the measuring device. In any case, a cantilever measures one of the states, $|0_e\rangle$ or $|1_e\rangle$. Certainly, the static displacement of the cantilever's tip is very small, and its measurement seems to be a complicated problem. That is why we would like to consider resonant methods of single-spin measurement using MRFM. 

However, an obstacle appears when one tries to apply conventional resonant methods for single spin detection using MRFM.
A single spin interacts with the cantilever as it does with any measuring device. A single spin interacting with any measuring device exhibits quantum jumps rather than smooth oscillations. Observation of resonant vibrations of a cantilever driven by harmonic oscillations of a single spin $z$-component contradicts to the principles of quantum mechanics. To generate resonant vibrations of the cantilever, we proposed a modification of the conventional MRFM technique \cite{bt1}. 
We use the fact that, under the action of a short{\it rf} pulse a spin must change its direction in a time-interval which is small in comparison with the cantilever's period of oscillations, $T_c=1/f_c$ (it can be a $\pi$-pulse or a pulse used in adiabatic inversion). During this short time interval, the single spin evolves as a pure quantum system in an external field. For the measuring device (cantilever), this evolution looks like a quantum jump. If one applies a periodic sequence of these short {\it rf} pulses with a period equal to $T_c/2$, one will observe resonant vibrations of the cantilever driven by periodic spin inversion. The cantilever will behave as though it is driven by an external harmonic force.

If one applies a periodic sequence of  $\pi$-pulses, the electron Rabi frequency of a $\pi$-pulse, $f_{eR}$, must be much larger than the resonant frequency of the cantilever: $f_{eR}\gg f_c$.
If one applies a periodic sequence of adiabatic inversions, the following condition must be satisfied: $f_c\ll |d{\vec B}_{eff}/dt|/2\pi B_1\ll f_{eR}$,
where ${\vec B}_{eff}$ is the effective field in the rotating reference frame.
(The second inequality is the standard adiabatic condition. The first inequality requires that the time of inversion must be much less than the the cantilever's period, $T_c$.)

While a proton single-spin measurement is possible using MRFM \cite{bt1}, it is much more difficult to detect a magnetic moment of a proton than the electron magnetic moment.
That is why we consider below using electron spin resonance (ESR) to detect a nuclear spin state.\\ \ \\
{\bf III.  An Electron Spin Resonance Strategy for Detecting a Nuclear Spin State}\\ \ \\

Suppose  we want to detect the state of a nuclear spin which interacts with an electron spin via the hyperfine interaction. In a large external magnetic field, $B_0$, the frequency of the electron spin transition depends on the state of a nuclear spin due to the hyperfine interaction. This dependence was discussed in \cite{kane} in connection with quantum computation using nuclear spins of $^{31}P$ atoms implanted in silicon. (See also \cite{kane1}-\cite{b2}.) Suppose that the frequency of the electron spin transition is $f_{e0}$ for the ground state of the nuclear spin, $|0_n\rangle$, and $f_{e1}$ for the excited state of the nuclear spin, $|1_n\rangle$. (See Fig. 2.) Then, in analogy with the scheme suggested in \cite{bt1}, if one applies a periodic sequence (with period, $T=T_c/2$) of short {\it rf} pulses with frequency $f_{e0}$, one will observe resonant vibrations of the cantilever only if the nuclear spin is in the ground state. Thus, one can measure the state of the nuclear spin using the quantum transitions of the electron spin.

We shall describe here an example of a nuclear spin quantum computer in which indirect single spin measurements could be made using a MRFM. Consider a chain of impurity paramagnetic atoms on the surface of a diamagnetic host. Assume that the nuclei of the host  have zero magnetic moments. Suppose that the distance between the impurity atoms is, $a=50\AA$, the distance between the ferromagnetic particle on the cantilever's tip (at equilibrium) and the target atom is, $d=100\AA$, the radius of the ferromagnetic particle is, $R=50\AA$ (see Fig. 1), and the energy levels for electron and nuclear spins of a paramagnetic atom are given in Fig. 2.
 
The magnetic field at the target atom, ${\vec B}_F$, caused by the ferromagnetic particle can be estimated as,
$$
{\vec B}_F={{\mu_0}\over{4\pi}}{{3{\vec{n}}({\vec m}{\vec n})-{\vec m}}\over{r^3}},\eqno(1)
$$
where $\mu_0=4\pi\times 10^{-7}$H/m is the permeability of free space, ${\vec n}$ in the unit vector which points from the center of the ferromagnetic particle to the atom, ${\vec m}$ is the magnetic moment of the ferromagnetic particle, $r$ is the distance between the atom and the center of the ferromagnetic particle. (The target atom is the atom in the sample which is closest to the ferromagnetic particle.) Substituting,
$$
m={{4}\over{3}}\pi R^3M,\eqno(2)
$$
where $M$ is the magnetization of the ferromagnetic particle, we obtain the magnetic field at the target atom produced by the ferromagnetic particle,
$$
B_{Fz}={{2}\over{3}}\mu_0M\Bigg({{R}\over{z}}\Bigg)^3,~B_{Fx}=B_{Fy}=0.\eqno(3)
$$
In (3), $z=R+d$ is the coordinate of the target atom. (We place the coordinate origin at the equilibrium position of the center of the ferromagnetic particle neglecting its static displacement.) Choosing $\mu_0M=2.2$T (which corresponds to an iron ferromagnetic particle), we get, $B_{Fz}=5.4\times 10^{-2}$T. The corresponding shift of the ESR frequency is ($\gamma_e/2\pi) B_{Fz}\approx 1.5$GHz. Here $\gamma_e$ is the electron gyromagnetic ratio ($\gamma_e/2\pi=2.8\times 10^{10}$Hz/T). Now, we  estimate the $z$-component of the magnetic field, $B^\prime_{Fz}$, at the neighboring impurity atom,
$$
B^\prime_{Fz}={{1}\over{3}}\mu_0 M\Bigg({{R}\over{r}}\Bigg)^3\Bigg[3\Bigg({{z}\over{r}}\Bigg)^2-1\Bigg].\eqno(4)
$$
Putting $r=\sqrt{z^2+a^2}$, we obtain $B^\prime_{fz}=3.6\times 10^{-2}$T. We assume here that the external magnetic field, $B_0$, is much greater than the field produced by the ferromagnetic particle. (Below we choose $B_0=10$T.) Thus, only the $z$-component,
$B^\prime_{Fz}$, is required to estimate the ESR frequency. The difference between the ESR frequencies for two neighboring atoms is,
$$
\Delta f^\prime_e=(\gamma_e/2\pi)(B_{Fz}-B^\prime_{Fz})\approx 500 MHz.\eqno(5)
$$
 The electron Rabi frequency, $f_{eR}$, which will drive a target electron spin must be smaller than $\Delta f^\prime_e$, to provide a selective measurement of the electron spin on the target atom.

Consider the cantilever's vibrations driven by the target electron spin. We take parameters of the cantilever from Ref. \cite{rzh}: the spring constant of the cantilever is $k_c=10^{-3}$N/m, the frequency of the cantilever is $f_c=1.4$kHz, the effective quality factor is $Q=10^3$, and the amplitude of thermal vibrations at room temperature ($T\approx 300$K) is $z_{rms}=5\AA$
($z_{rms}$ is the root-mean-square amplitude of thermal vibrations).
Assume that the temperature is $1$K. Scaling the square root dependence of $z_{rms}$ on $T$ \cite{bt1}, we obtain the characteristic amplitude of thermal vibrations of the cantilever \cite{rzh} at $T=1$K: $z_{rms}\approx 0.3\AA$.

The magnetostatic force on a target electron created by the ferromagnetic particle  can be estimated as,
$$
F_z=\pm \mu_B{{\partial B_{z}}\over{\partial z}},\eqno(6)
$$
for the positive and negative directions of the electron magnetic moment, respectively. Differentiating (3), we obtain the value of $F_z$,
$$
F_z=\mp 2\mu_B{{\mu_0M}\over{z}}\Bigg({{R}\over{z}}\Bigg)^3.\eqno(7)
$$
For $z=R+d$, the value of $F_z$ is: $F_z\approx\mp 10^{-16}$N. The magnetostatic force on the ferromagnetic particle is of the same magnitude, but it points in the opposite direction of the force on the target electron. When the magnetostatic force on the cantilever, $F_z(t)$, takes the values, $\pm F$, and the period of the function, $F_z(t)$, is equal to  the cantilever period, $T_c$, the stationary amplitude of the cantilever vibrations is \cite{bt1},
$$
z_c=4FQ/\pi k_c\approx 1.2\AA.\eqno(8)
$$
This amplitude of the stationary oscillations can be achieved after a transient process, whose  duration, $\tau_c$, can be estimated as,
$$
\tau_c=Q/\pi f_c\approx 0.2s.\eqno(9)
$$

Note that one does not have to ``wait'' until the stationary amplitude is reached. In fact, the experiment can be stopped when the {\it rms} amplitude of the driven cantilever's vibrations exceeds the thermal vibration amplitude, $z_{rms}\approx 0.3\AA$. This occurs when  the value of the driven amplitude, $z^\prime_c=\sqrt{2}z_{rms}\approx 0.4\AA$. Assuming the cantilever's amplitude increases as $z_c[1-\exp(-t/\tau_c)]$, we obtain the time of a single spin measurement:
$$
\tau_{m}=-\tau_c\ln(2/3) \approx 80ms.\eqno(10)
$$
The lifetime of the electron excited state must be greater than  $\tau_m$. (Otherwise a spontaneous electron transition can destroy the process of measurement.) 

Next, we shall estimate the deviation of the magnetic field at the target atom due to the cantilever's vibrations. For the value of $z^\prime_c\approx 0.4\AA$, the deviation of the magnetic field, $\Delta B_z$, is,
$$
\Delta B_z\approx\Bigg|{{\partial B_z}\over{\partial z}}\Bigg|z^\prime_c\approx 4\times 10^{-4}T.\eqno(11)
$$
The corresponding deviation of the ESR frequency is,
$$
\Delta f_e=(\gamma_e/2\pi)\Delta B_z\approx 10MHz.\eqno(12)
$$
To provide the ESR, the electron Rabi frequency, $f_{eR}$, must be greater than $\Delta f_e$. Thus, the requirement for the electron Rabi frequency, $f_{eR}$, can be written as,
$$
\Delta f_e\ll f_{eR}\ll\Delta f^\prime_e,\eqno(13)
$$
or $ 10MHz\ll f_{eR}\ll 500MHz$. Taking, for example, $f_{eR}=100$MHz we obtain the duration of an ``electron'' $\pi$-pulse: $\tau=1/2f_{eR}=5$ns. (We do not mention the condition $f_c\ll f_{eR}$, because in our case: $f_c\ll \Delta f_e$).
Certainly, to measure a nuclear state, the value of $f_{eR}$ must be less than the ESR hyperfine splitting, $2f_{hf}$. This imposes the requirement: $f_{hf}\gg\Delta f_e$. 
Note that the value of the electron Rabi frequency, $f_{eR}$, can be close to $\Delta f^\prime_{e}$ (or $2f_{hf}$ if $2f^\prime_{hf}<\Delta f_e$) if one uses the $2\pi k$-method. (See \cite{bct}-\cite{bdt}.) 

Finally, we estimate the dipole field produced by paramagnetic atoms. The main contribution to the dipole field is associated with the neighboring atoms. For any inner atom in the chain, two neighboring electron magnetic moments which point in the positive $z$-direction produce the magnetic field,
$$
B_{dz}=-2\mu_0\mu_B/4\pi a^3\approx -1.5\times 10^{-5}T.\eqno(14)
$$
The maximal contribution from all other paramagnetic atoms to the dipole field,  $ B^\prime_{dz}$,
does not exceed $-3\times 10^{-6}T$. (For a chain of 1000 paramagnetic atoms with electron spins in the ground state, the value of $B^\prime_{dz}$ at the center of the chain is: $B^\prime_{dz}\approx 0.202B_{dz}$.) The corresponding frequency shift of the ESR is,
$$
(\gamma_e/2\pi)|B_{dz}+B^\prime_{dz}|\le 500kHz.\eqno(15)
$$
This frequency shift is negligible compared to the estimated electron Rabi frequency, $f_{eR}\ge 100$MHz. Thus, to measure the  nuclear spin state of the target atom, one can use a periodic sequence of ``electron'' $\pi$-pulses with frequency,
$$
f\approx f_{e0}+(\gamma_e/2\pi)B_{Fz}.\eqno(16)
$$
\\ \ \\
{\bf IV. A Nuclear Spin Quantum Computer Using Single Spin Measurements}\\ \ \\ 
In this section, we assume that the MRFM can provide a single-electron spin measurement. We shall now discuss a nuclear spin quantum computer using this MRFM single-spin detection.

First, the single-spin MRFM can be used to create 100\%  polarization in a  nuclear spin chain. To do this, one should determine the initial state of each nuclear spin in the chain. (Note, that we assume  100\% polarization of electron spins. In the external magnetic field, $B_0\approx 10$T, at the temperature , $T\approx 1$K, the probability for an electron spin to occupy the upper energy level is approximately $\exp(-2\mu_BB_0/k_BT)\approx 1.4\times 10^{-6}$.  In fact, we can relax these restrictions on the magnetic field, $B_0$, or the temperature.) During the measurement process, one should use an even number of pulses to return the electron spin to the ground state. 

To create 100\%  polarization of the nuclear spins or to carry out a quantum computation, one must fix the $z$-coordinate of the ferromagnetic particle. This non-vibrating particle is not a classical measuring device. It is only a static source of the inhomogeneous magnetic field. One can imagine that the ferromagnetic particle could move along the spin chain. (See Fig. 3.) It cab be used to target each nuclear spin which is in the excited state.

According to (3) and (4), the target nuclear spin experiences an additional magnetic field, $B_{Fz}\approx 5.4\times 10^{-2}$T. The neighboring nuclear spin experiences an additional magnetic field, $B^\prime_{Fz}\approx 3.6\times 10^{-2}$T. The corresponding shifts of the NMR frequencies are:
$(\gamma_n/2\pi)B_{Fz}\approx 2.3$MHz and $(\gamma_n/2\pi)B^\prime_{Fz}\approx 21.5$MHz, where $\gamma_n$ is the nuclear gyromagnetic ratio. (Here and below we present estimations for a proton, $\gamma_n/2\pi\approx 4.3\times 10^7$Hz/T.) The frequency difference between the target nuclear spin and its neighbor is, $\Delta f^\prime_n\approx 800$kHz.

The Rabi frequency, $f_{nR}$, must be less than $\Delta f^\prime_n\approx 800$kHz to provide  a selective action of a $\pi$-pulse. Again, using the $2\pi k$-method \cite{bct}-\cite{bdt}, one can choose a value of $f_{nR}$ close to $\Delta f^\prime_n$. The corresponding duration of the $\pi$-pulse, $\tau=1/2f_{nR}$, must be greater than 630 ns.

The dipole field, $B_{dz}$, produced by two neighboring paramagnetic atoms (for inner nuclear spins in the chain) is given by (14). The maximal additional contribution of all other paramagnetic atoms is estimated to be: $B^\prime_{dz}\approx -3\times 10^{-6}$T. The corresponding shift of the NMR frequency is approximately 780 Hz, much less than the assumed value of the Rabi frequency, $f_{nR}\sim 100$kHz.
Thus, to drive the target nuclear spin into its ground state, one can apply a ``nuclear'' $\pi$-pulse with frequency $f\approx f_{n0}+(\gamma_n/2\pi)B_{Fz}$, where $f_{n0}$ is the NMR frequency of a paramagnetic atom with a ground state of the electron spin. (See Fig. 2.)
In this way, the whole chain of the nuclear spins can be initialized in its the ground state. In the same way, one can provide a one-qubit rotation for any selected nuclear spin.

Now we consider the possibility of implementing conditional logic in a chain of nuclear spins. The direct interaction between nuclear spins for inter-atomic distance, $a=50\AA$, is negligible. That is why, to provide conditional logic in a system of nuclear spins, we propose using an electron dipole field. Suppose that we want to implement a two-qubit quantum CONTROL-NOT (CN) logic gate. Because for our system, the ESR frequency, $f_{e0}+(\gamma_e/2\pi)B_{Fz}$, is the highest and the unique frequency in the chain, we will consider the ``inverse'' CN gate: the target qubit changes its state if the control qubit is in the ground state. (Here the target qubit is a nuclear spin which can change its state during the CN operation, but not necessarily the nuclear spin closest to the ferromagnetic particle.) 

Assume that the target qubit is any inner nuclear spin in the chain, and the control qubit is one of its neighboring nuclear spin. (See Fig. 4, where $A$ and $B$ are the control and target qubits.) We want to implement an ``inverse'' CN gate in three steps.

1. One sets the ferromagnetic particle near the control qubit (Fig 4a), and applies an ``electron'' $\pi$-pulse with frequency,
$$
f\approx f_{e0}+(\gamma_e/2\pi)B_{Fz}.\eqno(17)
$$
The electron Rabi frequency, $f_{eR}$, satisfies the inequalities: $f_{ER}< \Delta f^\prime_{e}$,
$2f_{hf}$. So, the electron magnetic moment of the left paramagnetic atom in Fig. 4 changes its direction only if the control qubit is in the ground state, $|0_n\rangle$. 

2. The ferromagnetic particle moves to the target qubit. (See Fig. 4b.) If the control qubit is in the excited state, than the electron magnetic moment of the left atom did not change its direction during the first step. In this case, the NMR frequency for the target qubit is,
$$
f_{n0}+(\gamma_n/2\pi)(B_{Fz}+B_{dz}+B^\prime_{dz}).\eqno(18)
$$
The third term in (18) is important for us: $(\gamma_n/2\pi)B_{dz}\approx 650$Hz. (The value $(\gamma_n/2\pi)B^\prime_{dz}$ depends on the position of the nuclear spin in the chain. We estimated that the range of variation for this term is approximately between 70 Hz and 130 Hz. The exact value of this term can be calculated or measured experimentally for each nuclear spin in the chain.) If the control qubit is in the ground state (as in Fig. 4) then the electron magnetic moment of the left atom changed its direction during the first step. In this case, the NMR frequency of the target qubit is,
$$
f_{n0}+(\gamma_n/2\pi)(B_{Fz}+B^\prime_{dz}),\eqno(19)
$$
because the dipole field produced by neighboring paramagnetic atoms cancels out: $B_{dz}=0$.

Next, one applies a ``nuclear'' $\pi$-pulse with frequency (19). The difference between the frequencies in (19) and (18) is approximately $650$Hz. Thus, the nuclear Rabi frequency, $f_{nR}$, must be less that 650 Hz. The corresponding duration of the ``nuclear'' $\pi$-pulse is $\tau>770\mu s$. (Using the $2\pi k$-method \cite{bct}-\cite{bdt}, one can choose the nuclear Rabi frequency, $f_{nR}$, close to  $650$Hz. In this case, one can neglect the last term $(\gamma_n/2\pi)B^\prime_{dz}$ in the expressions (18) and (19) because  $(\gamma_n/2\pi)B^\prime_{dz}\le 130$Hz.) Under the action of a ``nuclear'' $\pi$-pulse the target qubit changes its state if the control qubit is in its ground state.

3. To complete the CN gate, the ferromagnetic particle moves back to the control qubit. (See Fig. 4c.) Then one should again apply the ``electron'' $\pi$-pulse with frequency (17). This pulse drives the electron magnetic moment back to its ground state. A similar procedure can be applied if the target qubit is at either end of the  chain.

It is well-known that any quantum algorithm can be implemented with one-qubit rotations and two-qubit CN gates \cite{m7}. So, the MRFM designed for single spin measurement could accomplish three tasks: preparation of the spin system in the ground state, implementation of quantum computation, and measurement of the final state. Certainly, we require a long relaxation time for the nuclear spins in order to perform a considered quantum computation.\\ \ \\
{\bf V. A Nuclear Spin Quantum Computer Not Requiring a Single-Spin Measurement}\\ \ \\
In this section, we discuss a quantum computer design for a situation when a 
single-electron spin measurement in a chain of paramagnetic atoms is difficult to realize. 
In this case, following \cite{lloyd93,ber94}, we propose using an array of ordered non-interacting chains of nuclear spins of diamagnetic atoms,
$$
ABCABCABC...
$$
(See Fig. 5.) We also propose using an  elongated ferromagnetic probe which selects a whole column of spins, say the $B$-column. This cantilever produces a magnetic field, $B_F$, which is uniform in the direction of the column, as  shown in Fig. 5. 

In this case, one cannot use a MRFM to polarize the nuclear spins (unless a nuclear single-spin measurement will become possible). One has to rely on other methods to polarize  nuclei in solids. When the system is prepared in its ground state, one can use an elongated non-vibrating ferromagnetic particle to implement quantum computation. Note that one qubit is replaced by a column of identical qubits, as shown in Fig. 5. All qubits in the column will be in the same quantum state until the final measurement of the state.

To suppress unwanted dipole interactions between nuclear spins in a chain, one applies a large external magnetic field, $B_0$, at the angle, $\theta_0$ ($\cos\theta_0=1/\sqrt{3}$), to the direction of the chain. In this case, the stationary states differ only in the directions of individual spins (along the magnetic field). For any stationary state, the component of the dipole field along the direction of the magnetic field ($z$-component) is equal to zero. The influence of transversal components of the dipole field on the energy levels is negligible.  
So, we assume that the main interaction in the chain is the Ising interaction between  neighboring nuclear spins mediated by the chemical bonds (as in liquids, where the dipole coupling is averaged to zero by the rotational dynamics of molecules).

In this case, we can approximate the Hamiltonian of the chain by the Ising Hamiltonian,
$$
{\cal H}=-\sum_kf_kI^z_k-2\sum_kJ_{k,k+1}I^z_kI^z_{k+1},\eqno(20)
$$
where $f_k$ is the resonant frequency of the $k$-th spin without Ising interactions; $I^z_k$ is the operator of the $z$-component for spin 1/2; $J_{k,k+1}$ is the constant of Ising interaction in frequency units, and we set $2\pi\hbar=1$.

The resonant frequency of each spin depends on the state of its neighboring spins. For example, for the spin, $B$, we have the resonant frequency $f=f_B+J_{AB}+J_{BC}$ if both neighboring spins are in their ground states; $f=f_B+J_{AB}-J_{BC}$ if only the left ($A$) neighboring spin is in its ground state;  $f=f_B-J_{AB}+J_{BC}$ if only the right ($C$) neighboring spin is in its ground state; and  $f=f_B-J_{AB}-J_{BC}$ if both neighboring spins are in their excited states. The frequency $f_B$ equals $(\gamma_B/2\pi)B_0$, where $\gamma_B$ is the gyromagnetic ratio for the $B$-spins. Placing a non-vibrating ferromagnetic particle near the spin-column $B$, one changes the frequency $f_B$ to $f^\prime_B=f_B+(\gamma_B/2\pi)B_{Fz}$, where $(\gamma_B/2\pi)B_{Fz}\gg J$. 

To provide a one-qubit rotation for a selected $B$-spin column,  
one can apply an electromagnetic pulse with frequency, $f^\prime_B$, and with a Rabi frequency, $f_{BR}$: $J\ll f_{BR}\ll(\gamma_B/2\pi)B_{Fz}$.
 This pulse implements a one-qubit rotation independent of the states of neighboring spins and without significantly influencing any other $B$-spin or $A$-spin or $C$-spin.

To implement a CN gate, for example, with the $B$-spin as the target qubit
and with a left neighbor $A$-spin as a control qubit, one can apply in succession  two $\pi$-pulses with the Rabi frequency $f_{BR}<J$ and frequencies,
$$
f=f^\prime_B-J_{AB}+J_{BC},~f=f^\prime_B-J_{AB}-J_{BC}.\eqno(21)
$$
These two pulses change the state of the spin $B$ if the $A$-spin is in the excited state, independent of the state of the neighboring $C$-spin. (Again, the value of $f_{BR}$ can be chosen close to $J$ if one uses the $2\pi k$-method \cite{bct}-\cite{bdt}.)

At the end of the quantum computation using the quantum computer shown in Fig. 5, the vibrating cantilever can be used to measure the $z$-component of the nuclear spin averaged over the selected column, as in a conventional MRFM. The knowledge of the average spin  can be used to determine the result of the quantum computation \cite{m2}.  

Finally, if there is no way to prepare nuclear chains in their ground states, one can utilize the array shown in Fig. 5 as a statistical quantum computer similar to a liquid NMR quantum computer \cite{m1}-\cite{m6}. In this case, one operates with a statistical mixture of Ising chains. There are two advantages of the statistical quantum computer considered here compared with a statistical quantum computer based on liquid NMR. First, the proposed solid-state  design incorporates more qubits. Second, the sensitivity of the MRFM is higher than the sensitivity of the standard NMR measurements.\\ \ \\
{\bf Conclusion}\\ \ \\
In this paper, we proposed an MRFM  nuclear spin quantum computer which can operate at the temperature, $T\approx 1$K. We discussed procedures required for a single-spin measurement using MRFM. We pointed out an obstacle which prevents the application of a conventional MRFM to a single-spin measurement and described a way to overcome this obstacle. We also suggested a way to make an indirect measurement of the nuclear spin state via an electron spin and presented related estimates.

We have shown that a single-spin MRFM can provide everything that is needed for a nuclear spin quantum computer: polarization of nuclear spins, quantum transformations, and the final measurement. If a single-spin measurement is too difficult or not possible, we proposed the use of an array of nuclear spin chains. In this case, one can implement quantum computation  and measure the average spin in each column of a two-dimensional array. If there is no way to initialize nuclear chains in their ground states, this array could be used as  a statistical quantum computer which  features more qubits than a liquid NMR quantum computer.\\ \ \\
{\bf Acknowledgments}\\ \ \\
This work  was supported by the Department of Energy under contract W-7405-ENG-36 and by the National Security Agency.
\newpage
\quad\\
{\bf Figure captions}\\ \ \\
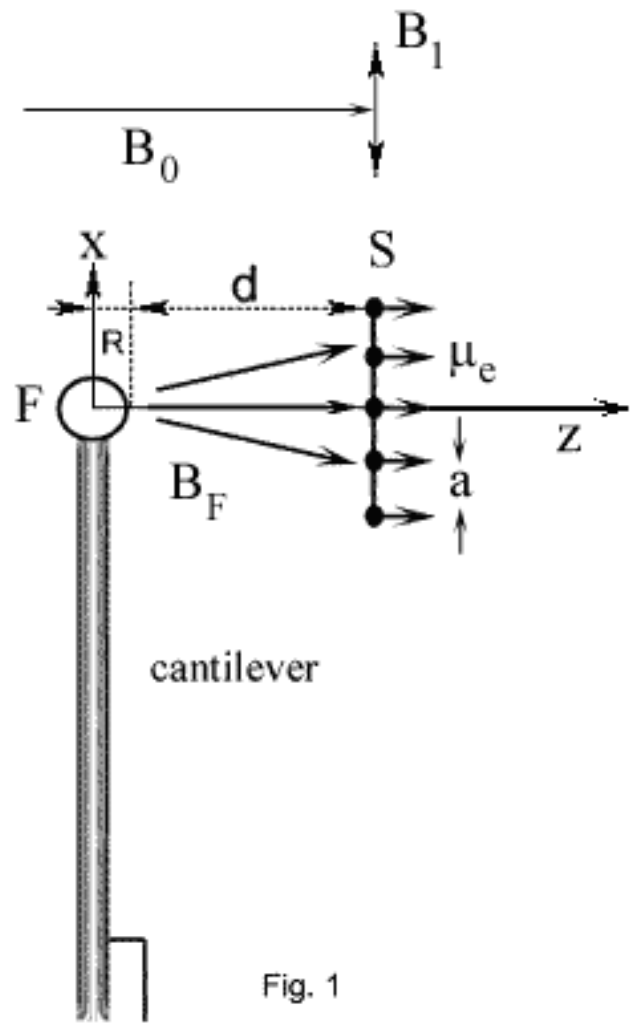
Fig. 1:~ A nuclear spin quantum computer based on the MRFM using single spin measurement; ${ B}_0$ is the permanent magnetic field; ${ B}_1$ is the radio-frequency magnetic field; ${\vec B}_F$ is the non-uniform  magnetic field produced by a ferromagnetic particle, $F$, in the sample, $S$; $R$ is the radius of the ferromagnetic particle; $d$ is the distance between the ferromagnetic particle and the targeted atoms; $a$ is the distance between  neighboring impurity atoms. The origin is placed at the equilibrium position of the center of the ferromagnetic particle. Arrows on the sample, $S$, show the direction of electron magnetic moments, $\mu_e$.\\ \ \\
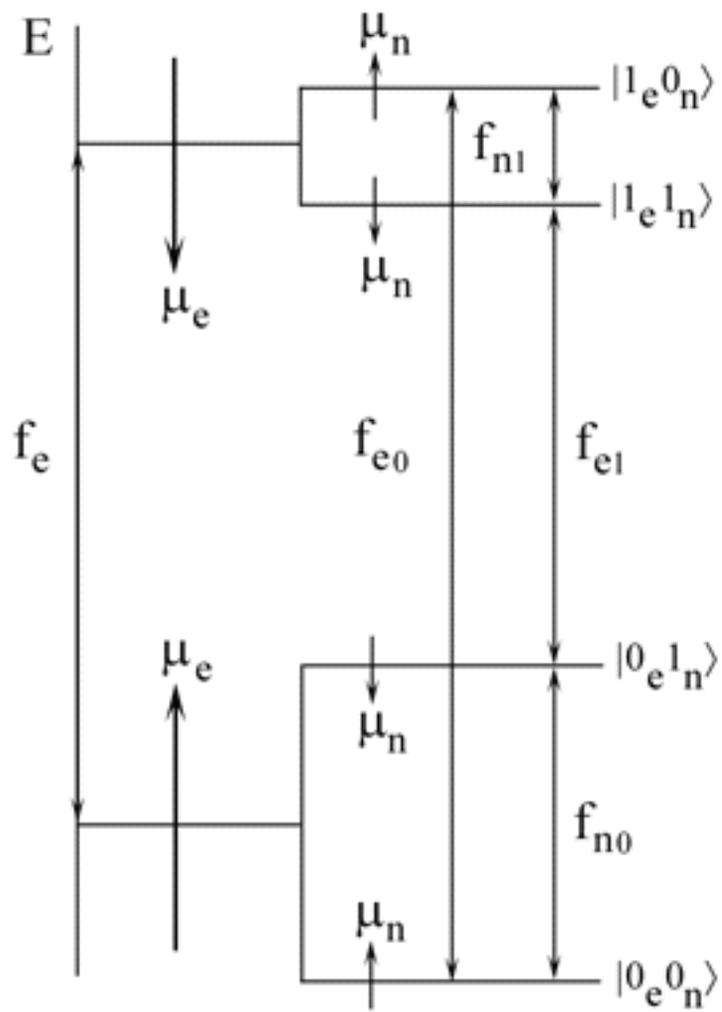
Fig. 2:~ Energy levels for electron and nuclear spins 1/2 of a paramagnetic atom in a high external magnetic field. The electron and nuclear magnetic moments are indicated by $\mu_e$ and $\mu_n$. $f_e$ and $f_n$ are the frequencies of the electron spin resonance (ESR) and nuclear magnetic resonance (NMR) in the external magnetic field without the hyperfine interaction. The frequencies $f_{e0}=f_e+f_{hf}$, $f_{e1}=f_e-f_{hf}$, $f_{n0}=f_n+f_{hf}$, $f_{n1}=f_{hf}-f_n$.
$f_{hf}$ is the hyperfine frequency (half of the hyperfine splitting of the ESR). $f_{eq}$ is the ESR frequency for the nuclear state $|q_n\rangle$ ($q=0,1$). $f_{nq}$ is the NMR frequency for the electron state $|q_e\rangle$. We assume that the magnetic moment of the nucleus and the hyperfine constant are positive, and $f_{hf}>f_n$. In the external field, $B_0=10$T, $f_e\approx 280$GHz and $f_n\approx 430$MHz (for a proton). \\ \ \\
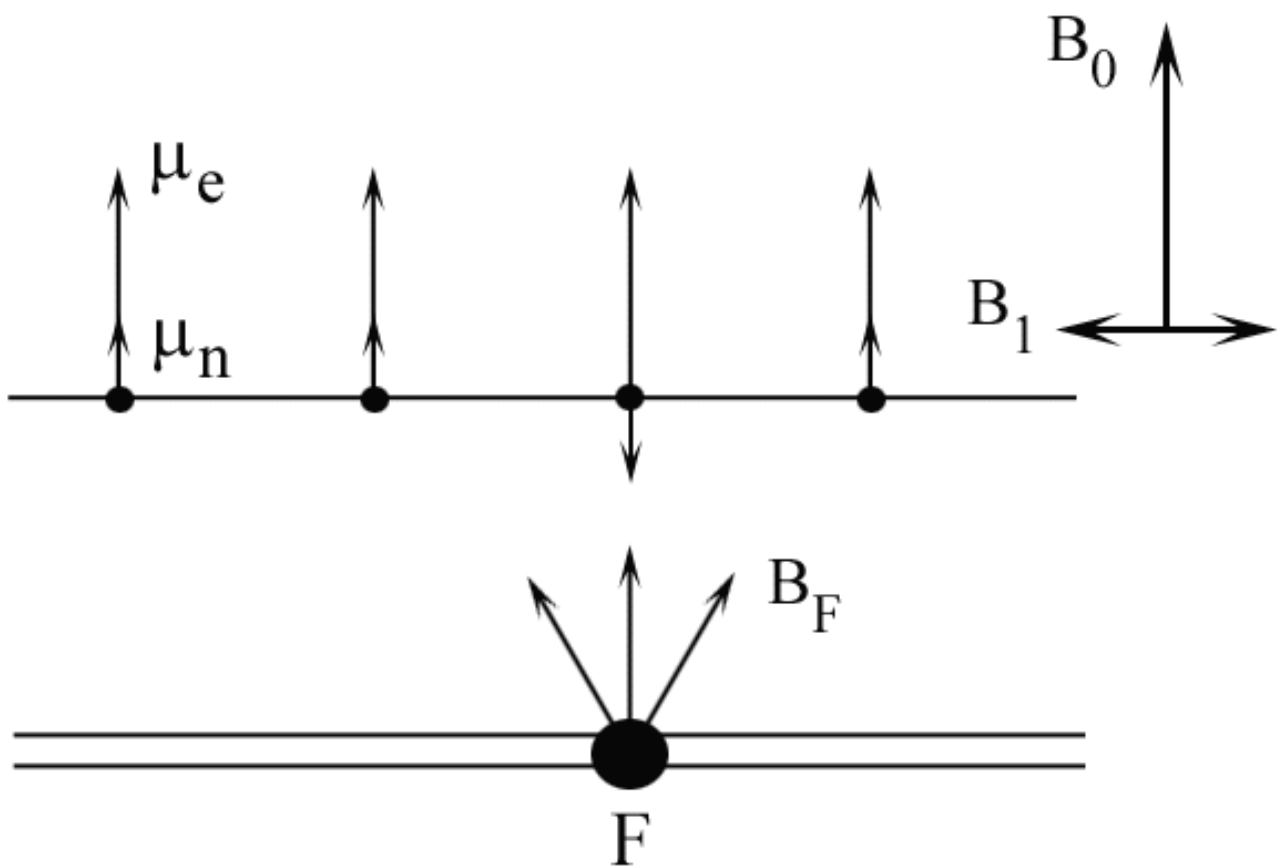
Fig. 3:~ The polarization of nuclear spins: The non-vibrating ferromagnetic particle targets nuclear spins which are initially in the excited state. \\ \ \\
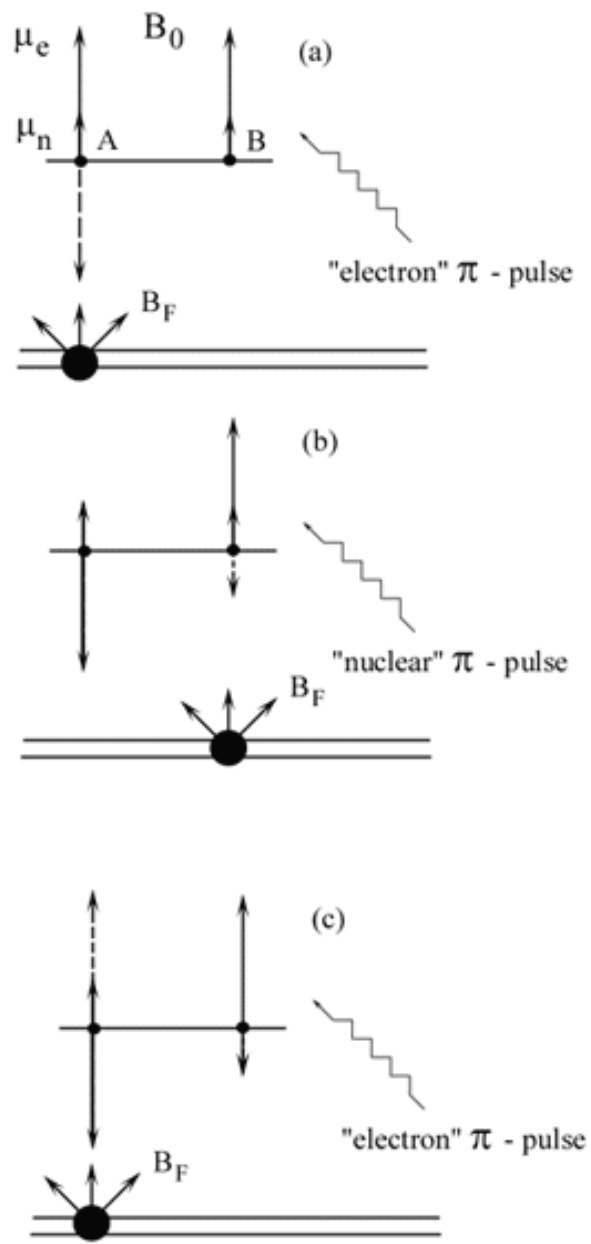
Fig. 4:~ Implementation of a quantum CN gate: (a) an ``electron'' $\pi$-pulse drives the electron magnetic moment of the control qubit (nuclear spin $A$) if $A$  is in the ground state; (b) a ``nuclear'' $\pi$-pulse causes a transition in the target qubit (nuclear spin $B$) if the control qubit $A$ is in the ground state; (c) ``electron'' $\pi$-pulse drives the electron magnetic moment back into the ground state.\\ \ \\
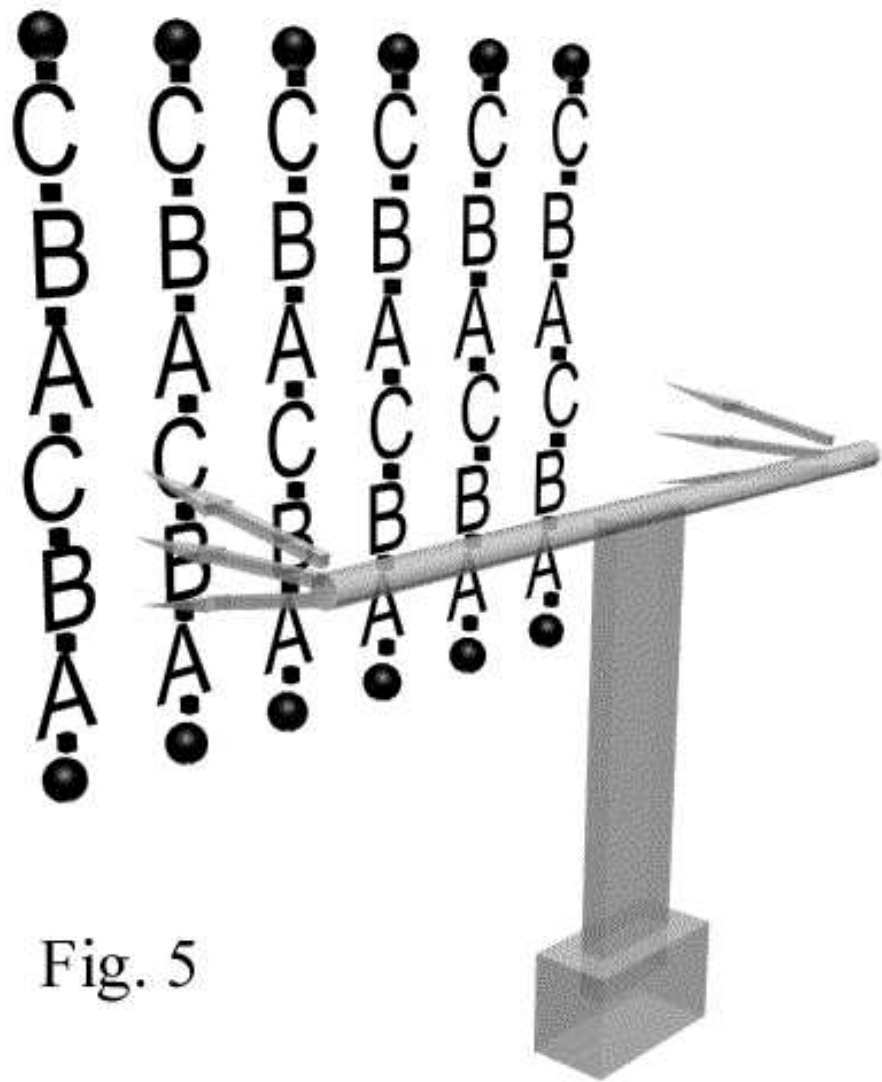
Fig. 5:~ A quantum computer based on an array of ordered non-interacting nuclear spin chains using MRFM.
\newpage

\end{document}